\documentstyle[12pt,aps,here,bezier]{revtex}
\tightenlines
\newcommand{\be}{\begin{equation}}
\newcommand{\ee}{\end{equation}}
\newcommand{\bee}{\begin{eqnarray}}
\newcommand{\eee}{\end{eqnarray}}
\newcommand{\sect}[1]{\section{#1}}

\begin{document}
\author{C.~von~Ferber$^{1}$ and Yu.~Holovatch$^{2}$}
\address{
$^1$Fachbereich Physik, Universit\"at - Gesamthochschule - Essen,\\
D-45117 Essen, Germany \\
$^2$Institute for Condensed Matter Physics,
Ukrainian Academy of Sciences,\\
UA-290011 Lviv, Ukraine
}
\title{Polymer stars in three dimensions. Three loop results}
\maketitle
\begin{abstract}
We study scaling properties of self avoiding polymer stars and
networks of arbitrary given but fixed topology. We use massive field
theoretical renormalization group framework to calculate critical
exponents governing their universal properties (star exponents).
Calculations are performed directly in three dimensions,
renormalization group functions are obtained in three loop
approximation. Resulting asymptotic series for star exponents are
resummed with the help of Pad\'e-Borel and conformal mapping
transformation.
\end{abstract}
%
\sect{Introduction \label{sect:i}}

Many peculiarities in the behavior of polymer chains immersed in
a good solvent are well understood qualitatively and described with
high precision on the quantitative level due to the application of
renormalization group methods \cite{BreLeGZin76,Zin89,Ami78}
originating from
field theory \cite{BogShir59}. This progress was initiated by
De~Gennes ideas linking the model of the
$d$-dimensional ferromagnet of $O(n)$-symmetry near its critical
temperature $T=T_{c}$  to self-avoiding walks
({\it SAW}) on a lattice of equal dimensionality $d$
\cite{DeGen72,DeGen79}. In particular correlation functions of the
$O(n)$-symmetric model  correspond to those of {\it SAW}s in the limit $n=0$.
 Let us recall that the critical exponents
$\nu$ and $\gamma$ describing the behavior of the
correlation length ($\xi$) and magnetic susceptibility ($\chi$)
of the $O(n)$-model in the vicinity of critical temperature $T_{c}$,
$$
\xi \sim \tau^{-\nu},
$$
$$
\chi \sim \tau^{-\gamma}, \quad \tau = \frac{T-T_{c}}{T_{c}}.
$$
In polymer theory these exponents correspond to size and configuration
number exponents
$\nu$ and $\gamma$ describing the average square end-to-end distance
$R^{2}$
\be
\label{eq:i3}
R^{2} \sim N^{2\nu},\quad  N \gg 1
\ee
of a single chain of $N$ monomers and the number ${\cal Z}_{N}$ of
possible ways of realization of a $SAW$ of $N$ steps on a given lattice:
\be
\label{eq:i4}
{\cal Z}_{N} \sim \mu^{N} N^{\gamma -1}, \quad N \gg 1,
\ee
where $\mu$ is a non-universal fugacity.

These laws have been generalized to describe polymers linked at their
endpoints to form polymer networks
\cite{DesCloiz80,Dupl86,DuplSal86,Dupl87,Dupl89}.
It has been shown that for a given network $G$ (\ref{eq:i4}) holds
with an exponent $\gamma=\gamma_{G}$ characteristic of the network.
Furthermore a subset of networks called polymer stars consisting of
$F$ polymer chains linked at one endpoint gives rise to a basic
series of exponents $\gamma_F$. Just as any network may be disassembled
to a number of stars cutting polymer chains, its characteristic
exponent $\gamma_{G}$ is expressed in terms of the exponents
$\gamma_F$ of its constituent stars.
Whereas the exponents (\ref{eq:i3}),(\ref{eq:i4}) were intensively studied
within the renormalization group approach to high order and by different
methods, this is not the case for the star exponents.
Their counterparts in the $O(n)$-symmetric model were introduced
as anomalous dimensions of certain composite operators \cite{WalZia75}
but with no direct physical interpretation.
In polymer theory they were introduced independently \cite{DesCloiz80,Dupl86}
to characterize networks and contacts between chains and later linked
to the $n=0$ limit of the composite operators \cite{SchaFer92}.
With a number of Monte Carlo simulations at hand
\cite{BatKre89,BarTre87,Grassberger} they form an attractive and
interesting  field for analytic calculations. In this article we present our
results for star exponents obtained in the frames of massive field
theory renormalization scheme directly in three dimensions. This
scheme of calculations as initiated by Parisi \cite{Par80}
avoids the $\varepsilon$-expansion and thus forms an independent check for
the validity of previous results
\cite{BatKre89,BarTre87,DuplFer92,SchaFer92}. As to our knowledge
it is applied here in the context of polymer star properties for the
first time.

\begin{figure} [H]
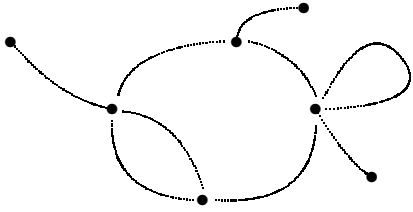

\begin{center}
\input net.pic
\end{center}
\caption{
A polymer network $G$. It is characterized by
the numbers $n_{F}$ of
$F$-leg vertices. Here $n_{1}=3$, $n_{3}=2$,
$n_{4}=1$, $n_5=1$.
}
\label{fig:net}
\end{figure}

\begin{figure} [H]
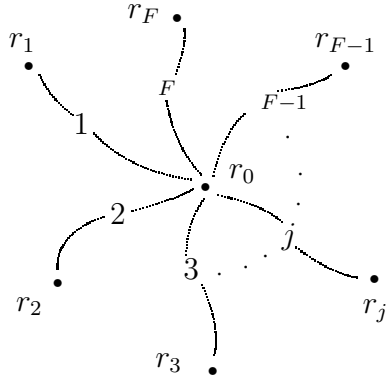

\begin{center}
\input star.pic
\end{center}
\caption{
Star polymer.
}
\label{fig:star}
\end{figure}

\begin{figure} [H]
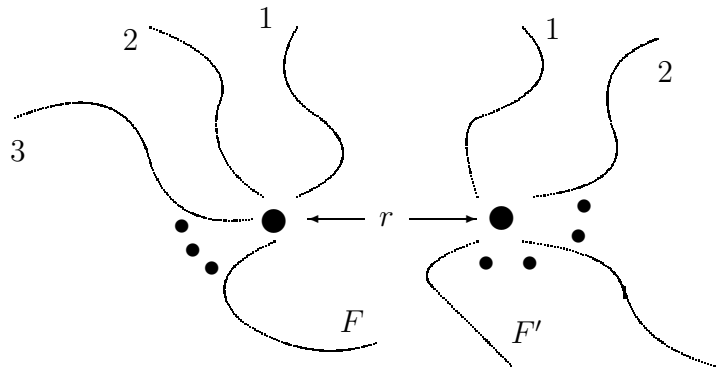

\begin{center}
\input star2.pic
\end{center}
\caption{
Two star polymers of functionalities $F$ and $F'$ at a
distance~$r$.
}
\label{fig:star2}
\end{figure}

Let us give a short account of the setup of the article.
The next section introduces the model and defines several series of
star exponents giving  scaling relations between them.
In section 3 we calculate the renormalization group functions as
series in the coupling to 3rd loop order . Being
asymptotic these series are resummed in section 4 by means of
Pad\'e-Borel and conformal mapping technique, we give the
numerical values of star exponents and discuss them together
with results of other approaches.
In appendices we show the correspondence between the loop integrals
entering the theory and their Feynman diagrams, and give the
details of the pseudo-$\varepsilon$-expansion for the star
exponents.
%
\sect{Polymer networks, star polymers and star exponents
\label{sect:p}}
Consider a polymer network consisting of long polymer chains tied
together at their endpoints (fig. \ref{fig:net}).
Mapping such a network on an equivalent field theory one may prove
\footnote{ This proof is valid for polymer networks of arbitrary
architecture at $d<4$ and for a large class of molecular weight
distributions of the strands.
}
\cite{DuplFer92,SchaFer92} that as in the case of a linear
chain \cite{DesCloiz75,SchafWit80} this theory is multiplicatively
renormalizable. The scaling properties of the network are determined
by its  "star-like" vertices, connecting the extremities of the
chains.  Let us consider a single star polymer (fig. \ref{fig:star})
with $F$ legs of $N$ monomers each. It can be shown that for long
chains $N\to\infty$ the number of self-avoiding configurations will
scale according to \cite{Dupl89} :
\be
\label{eq:p1}
{\cal Z}_{N,F} \sim \mu^{FN}
N^{\gamma_{F}-1}, \quad N \gg 1.
\ee
Formula (\ref{eq:p1}) can be considered as the generalization of
(\ref{eq:i4}), $\mu$ still being the non-universal connectivity
constant and values $\gamma_{F}$ give us the first example of star
exponents. In the case $F=1,2$ one still has the single polymer chain:
$\gamma_{1}=\gamma_{2}=\gamma$ and for $F\geq3$ $\gamma_{F}$ form a
set of independent critical exponents. For the two-dimensional case
the values of $\gamma_{F}$ are known exactly \cite{Dupl89,DuplSal86}:
$$
\gamma_{F} = [68 + 9F(3-F)]/64.
$$
Exponents $\gamma_F$ give an example of geometrical exponents
associated with special vertices where infinite critical objects are
fused together. As mentioned in the introduction there is no
direct analogy to critical exponents describing the 2nd
order phase transition in $O(n)$-symmetric model (see, however
formulas  (\ref{eq:p10}),(\ref{eq:rg15}) of this article).
Let us note here a similar problem with an {\it infinite} spectrum of
scaling dimensions.
Consider a $d=2$ dimensional percolation cluster and the probability
$P_F$ of a pinching point joining $F$ peninsulas of the infinite
incipient cluster. Near the percolation
threshold $p_c$ this scales like \cite{DuplSal87,Dupl90}:
$$
P_F \sim (p-p_c)^{\beta_F},
$$
where $\beta_F$ is another basic set of geometrical critical exponents
related to $\gamma_F$ in $d=2$.
Another
example are the so-called contact exponents introduced in \cite{Dupl89}
(see \cite{DesCloiz80}  as well) to describe the mutual behavior of
two polymer stars having $F$ and $F'$ legs as function of the distance
$r$ of their cores.  (fig.
\ref{fig:star2}).
Due to the hard-core repulsion the  probability of approaching the
cores of these two polymers at distance $r$ vanishes as
$$
P(r) \sim r^{\theta_{F,F'}}, \quad  r \rightarrow 0.
$$
Again $\theta_{F,F'}$ form a set of universal contact exponents
which can not be expressed in terms of $\nu$ and $\eta$ (\ref{eq:i3}),
(\ref{eq:i4}).  Let us note that there exist scaling relations between
different sets of star exponents. Choosing $\gamma_F$ exponents as the
"basic" ones one gets for the contact exponents $\theta_{F,F'}$
\cite{Dupl89}:
$$
\nu\theta_{F,F'} -1 = \gamma_{F} + \gamma_{F'} - \gamma_{F+F'}.
$$
Considering a general network $G$ of arbitrary but fixed topology,
made of $F$ chains of equal length $N$ and tied together in vertices (see
fig. \ref{fig:net}) one has for the asymptotic number of self-avoiding
configurations ${\cal Z}_{G}$:
$$
{\cal Z}_{G} \sim \mu^{F N} N^{\gamma_{G}-1}, \quad N \gg 1,
$$
where the expression for the critical exponent $\gamma_{G}$ reads:
$$
\gamma_{G} - 1 +\nu d =  \sum _{F\geq1}n_{F}
\Big[(\gamma_F-1+\nu d) - \frac{F}{2}(\gamma_2-1+\nu d)\Big],
$$
here  $n_{F}$ is the number of vertices with $F$ legs.

Thus knowing one set of exponents for single  star polymers one can
obtain geometrical exponents characterizing any polymer network
\cite{Dupl86,DuplSal86,DuplFer92,SchaFer92}.
So in what follows below we will consider mainly the properties of a
single polymer star immersed in a good solvent with arbitrary but
fixed number of arms (see fig. \ref{fig:star}).
We are using the continuous chain model introduced by Edwards
(see e.g. \cite{DesCloizJan87}). Consider an ensemble of $F$
disconnected branches and put in correspondence to each branch $a$ a
path ${\bf r}_{a}(s)$, parametrized by
$0 \leq s \leq S_{a}, a=1,2,...,F$ ($S_{a}$
being the Gaussian surface of $a$th branch). One
can describe such a system by the following Hamiltonian:
\bee
\frac{1}{k_{B}T}H_{F} &=& \frac{1}{2} \sum_{a=1}^{F}
\int\limits_{0}^{S_{a}}ds (\frac{d{\bf r}(s)}{ds})^{2} + \nonumber \\
&&\frac{u_{0}}{2} \sum_{a,b=1}^{F}
\int\limits_{0}^{S_{a}}ds \int\limits_{0}^{S_{b}}ds'
\delta^{d}({\bf r}_{a}(s) - {\bf r}_{b}(s')).  \nonumber
\eee
The partition function is obtained as a functional integral over
all possible configurations of polymer system divided by its volume
$\Omega$ thus dividing out identical configurations just translated in
space:
\be\label{eq:p5}
{\cal Z}\{S_{a}\} = \frac {1}{\Omega} \int D[{\bf r}_{1},...,
{\bf r}_{F}] \exp [-\frac {H_{F}}{k_{B}T}],
\ee
here the symbol $D[{\bf r}_{1},...,{\bf r}_{f}] $
includes normalization such that ${\cal Z} \{S_{a}\} = 1$ for
$u_{0}=0$.  To make the exponential of $\delta$-functions and the
functional integral well defined in bare theory a cutoff $s_0$ has to
be introduced such that all simultaneous integrals of any variables
$s$ and $s'$ are cut off by $|s-s'| > s_0$.
With this we can relate the Gaussian surface $S$ of a path to the
notion of steps $N$ of a self avoiding walk by
$$
    N = S/s_0.
$$
Introducing into (\ref{eq:p5}) product of $\delta$-functions
$$
\prod_{a=2}^{F} \delta^{d}({\bf r}^{a}(0)-{\bf r}^{1}(0))
$$
ensuring  the ``star-like" configuration of a set of $F$ chains
(c.f. fig. \ref{fig:star}), one gets for the partition function of the
polymer star:
\bee
{\cal Z}_F\{S_{a}\} \equiv  {\cal Z}_F\{S_{1},...,S_{F}\} & = &
\frac {1}{\Omega} \int D[{\bf r}_{1},...,{\bf r}_{F}]
\exp [-\frac {H_{F}}{k_{B}T}] \times
\nonumber \\
&&\prod_{a=2}^{F} \delta^{d}({\bf r}^{a}(0)-{\bf r}^{1}(0)).
\nonumber
\eee

Besides the partition functions one can define correlation   functions
of an $F$-arm star. For the core-endpoint correlation function one has
\cite{Dupl86,DuplSal86,DuplFer92,SchaFer92}:
\be\label{eq:p6a}
\hat{P}_{F}({\bf r}_{0};{\bf r}_{1},...,{\bf r}_{F}; S_{1},...,S_{F})=
\langle \prod_{a=1}^{F} \delta^{d}({\bf r}^{a}(0)-{\bf r}_{0})
\delta^{d}({\bf r}^{a}(S_{a})-{\bf r}_{a}) \rangle,
\ee
here ${\bf r}_{0}$ is the coordinate of the core,
${\bf r}_{1},...,{\bf r}_{F}$ being coordinates of chain
endpoints, $S_{1},...,S_{F}$ being their Gaussian
surfaces. The averaging in (\ref{eq:p6a}) means:
$$
\langle...\rangle=\frac { \frac {1}{\Omega}
\int D[{\bf r}_{1},...,{\bf r}_{F}]
\exp [-\frac {H_{F}}{k_{B}T}] (...)} {{\cal Z}_{F}\{S_{a}\}}.
$$
For the Green functions one has:
$$
(2 \pi)^{d}  \delta ^{d}[{\bf p}_{0} + \sum_{a=1}^{F}{\bf p}_{a}]
G_{F}({\bf p}_{0};{\bf p}_{1},...,{\bf p}_{F};S_{1},...,S_{F})=
$$
\be\label{eq:p6b}
{\cal Z}_{F}(S_{1},...,S_{F}) \int \prod _{a=0}^{F} [d^{d} {\bf r}_{a}
\exp [i{\bf p}_{a}{\bf r}_{a}]
\hat{P}_{F}({\bf r}_{0};{\bf r}_{1},...,{\bf r}_{F};S_{1},...,S_{F}).
\ee
The mapping to field theory is performed by a Laplace transformation
from the Gaussian surfaces $S_{a}$ to conjugated chemical potential
variables ("mass variables") $\mu _{a}$:
\be\label{eq:p7}
\tilde{{\cal Z}}\{\mu _{a}\} = \int \prod_{b=1}^{F}dS_{b} \exp[-\mu_{b}
S_{b}] {\cal Z}_{F} \{S_{a}\}.
\ee
As a result an ensemble of $F$ polymer chains is described by a
field theory with the Lagrangean ${\cal L} \{\phi_{a}, \mu_{a}\}$
involving a separate field $\vec{\phi}_{a}\equiv \vec{\phi}_{a}({\bf r})$
for each chain:
\be\label{eq:p8}
{\cal L} \{\phi_{b}, \mu_{b}\} = \frac{1}{2} \sum_{a=1}^{F} \int
d^{d}{\bf r}(\mu_{a} |\vec{\phi}_{a}|^{2} + \frac {1} {2}
|\nabla \vec {\phi}_{a}|^{2}) + \frac {u_{0}}{8} \sum_{a,a'=1}^{F}
\int d^{d}{\bf r}|\vec{\phi}_{a}|^{2}|\vec{\phi}_{a'}|^{2} ,
\ee
here $\vec{\phi}_{a}$ is an $n$-component vector field
$\vec{\phi}_{a}= (\phi^{1},...,\phi^{n})$,
$|\vec{\phi}_{a}|^{2}= \sum_{\alpha=1}^{n} ({\phi}_{a}^{\alpha})^{2}$.
However, the perturbation expansion of field theory (\ref{eq:p8})
at arbitrary value of $n$ results in particular in some diagrams,
which do not appear in polymer theory . As it is well known
\cite{DeGen72,DeGen79} such diagrams (involving closed loops of
propagator lines connected to the remainder of the diagram by
interaction lines only) can be suppressed by taking the limit $n=0$.

Returning now to the partition function of a single polymer
star ${\cal Z}_{F}\{S_{b}\}$ and using the Laplace transformation
(\ref{eq:p7}) we can represent its Laplace transform
$\tilde{{\cal Z}}_{F}\{\mu_{b}\}$ in the following form:
\bee
\tilde{{\cal Z}}_{F}\{\mu _{b}\}
&=& \frac {1}{\Omega} \int \ d^{d}{\bf r}_{0}
\prod_{a=1}^{F} d^{d}{\bf r}_{a}
\int D[\phi _{a}(r)] \times  \nonumber \\
&& \prod_{a=1}^{F} \phi_{a}(r_{0})
\phi_{a} (r_{a}) \exp[-{\cal L} \{\phi_{b},\mu_{b}\}] \mid_{n=0} .
\nonumber
\eee
In field theory the star vertex is  related to the local composite
operator $\prod_{a=1}^F\phi_a$ of $F$ distinct zero component fields.
Formally this is the $n=0$ limit of an operator known in
$n$-component field theory \cite{WalZia75}:
\be\label{eq:p10}
\sum_{\alpha_{1},...,\alpha_{F}=1}^{n}
N^{\alpha_{1},...,\alpha_{F}}
\phi^{\alpha_{1}}(r)....\phi^{\alpha_{F}}(r) ,
\ee
where $N^{\alpha_{1},...,\alpha_{F}}$
is a zero trace symmetric $SO(n)$ tensor:
$$
\sum_{\alpha=1}^{n}
N^{\alpha,\alpha,\alpha_{3},...,\alpha_{F}} = 0.
$$

Now it can be checked diagrammatically that Green functions
involving an operator with $N^{\alpha_{1},...,\alpha_{F}}$ symmetry
coupled to $F$ external fields $\phi_{a}^{\alpha}(r_{a})$ in the
limit $n=0$ coincide with Green functions (\ref{eq:p6b}) of the
$F$-arm star polymer. Thus in order to find star exponents in the
field-theoretical renormalization group program one may consider the
problem of renormalization of composite operators (\ref{eq:p10}).
This will be the subject of the subsequent section.
%
\sect{Renormalization group functions in three loop approximation
\label{sect:rg}
}
As was shown in the previous section considering the behavior
of the star-like vertices in polymer theory one faces the problem
of renormalization of a field theory containing two couplings
one being of $O(n)$ symmetry and described by tensor
$S_{\alpha_{1},...,\alpha_{4}}$, the other being of traceless $SO(n)$
symmetry ($N^{\alpha_{1}.,,,.\alpha_{F}}$):
\bee
&&{\cal L} \{\phi_{b}, \mu_{b}\} = \frac{1}{2}  \int
d^{d}{\bf r}[(\mu_{0}^{2} |\vec{\phi}|^{2} +
|\nabla \vec {\phi}|^{2}) +
\frac {u_{0}}{4!}
S_{\alpha_{1},...,\alpha_{4}}
\phi^{\alpha_{1}}....\phi^{\alpha_{4}}+ \nonumber \\
&&\frac{v_{0}}{4!}
N^{\alpha_{1},...,\alpha_{F}}
\phi^{\alpha_{1}}....\phi^{\alpha_{F}}].
\label{eq:rg1}
\eee
Let us point out that the problem of composite
field renormalization in the case of traceless symmetry we
are going to tackle here was considered in the frames of
$\varepsilon$-expansion in \cite{WalZia75} in the field-theory
context, in polymer context $\varepsilon$-expansion results up to
$\varepsilon^{3}$-order were analysed in  \cite{DuplFer92,SchaFer92}.
We will return to these results later. The distinct feature of our
study is that we are going to apply here equations of the massive
field theory. As is well known in order to analyze the
expressions appearing in such an approach one can either apply
$\varepsilon$-expansion (and thus re-derive the results of
\cite{DuplFer92,SchaFer92,WalZia75}) or as it was proposed by Parisi
\cite{Par80} consider renormalization group functions directly in
the dimension of space of interest (here we are interested in the
three-dimensional case). So performing calculations in the spirit of
\cite{Par80} enables us to proceed directly in three dimensions.

\begin{figure} [H]
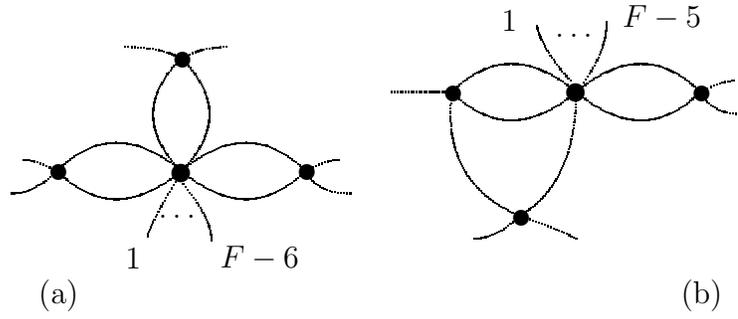

\begin{center}
\input stargr.pic
\end{center}
\caption{
Additional graphs appearing in the function $\Gamma^{F}$ in
three-loop approximation.
}
\label{fig:stargr}
\end{figure}

\begin{figure} [H]
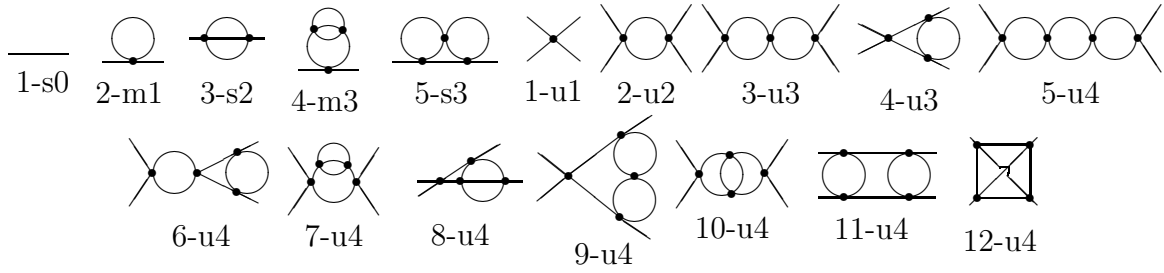

\begin{center}
$\begin{array}{c}\input{1-s0.pic}\\ \mbox{1-s0}
\end{array}$
$\begin{array}{c}\input{2-m1.pic}\\ \mbox{2-m1}
\end{array}$
$\begin{array}{c}\input{3-s2.pic}\\ \mbox{3-s2}
\end{array}$
$\begin{array}{c}\input{4-m3.pic}\\ \mbox{4-m3}
\end{array}$
$\begin{array}{c}\input{5-s3.pic}\\ \mbox{5-s3}
\end{array}$
$\begin{array}{c}\input{1-u1.pic}\\ \mbox{1-u1}
\end{array}$
$\begin{array}{c}\input{2-u2.pic}\\ \mbox{2-u2}
\end{array}$
$\begin{array}{c}\input{3-u3.pic}\\ \mbox{3-u3}
\end{array}$
$\begin{array}{c}\input{4-u3.pic}\\ \mbox{4-u3}
\end{array}$
$\begin{array}{c}\input{5-u4.pic}\\ \mbox{5-u4}
\end{array}$
$\begin{array}{c}\input{6-u4.pic}\\ \mbox{6-u4}
\end{array}$
$\begin{array}{c}\input{7-u4.pic}\\ \mbox{7-u4}
\end{array}$
$\begin{array}{c}\input{8-u4.pic}\\ \mbox{8-u4}
\end{array}$
$\begin{array}{c}\input{9-u4.pic}\\ \mbox{9-u4}
\end{array}$
$\begin{array}{c}\input{10-u4.pic}\\ \mbox{10-u4}
\end{array}$
$\begin{array}{c}\input{11-u4.pic}\\ \mbox{11-u4}
\end{array}$
$\begin{array}{c}\input{12-u4.pic}\\ \mbox{12-u4}
\end{array}$
\end{center}
\caption{
Graphs of functions $\Gamma^{(2)}$, $\Gamma^{(4)}$ in
three-loop approximation.
}
\label{fig:graphs}
\end{figure}

The critical properties of the field theory (\ref{eq:rg1}) can be
extracted from the coefficients $\beta_{u}, \beta_{v}, \gamma_{\phi},
\gamma_{\phi^{2}}$ of the Callan-Symanzik equation for the renormalized
$N$-point vertex function $\Gamma_{R}^{(N)}$ (see e.g.
\cite{Ami78,BreLeGZin76,Zin89}).  As far as we are interested in the
renormalization group functions and their derivatives over the
coupling constants at the Heisenberg fixed point determined
by the stable solution of $\beta_u(u^{*}\neq 0, v^{*}=0)=0$ we need
to calculate vertex functions of the symmetric interaction
$S_{\alpha_{1},...,\alpha_{4}} \phi^{\alpha_{1}}....\phi^{\alpha_{4}}$
$\Gamma^{(2)}, \Gamma^{(4)}$ together with $F$-point vertex function
$\Gamma^{(F)}$ with only one
$N^{\alpha_{1},...,\alpha_{F}}\phi^{\alpha_{1}}....\phi^{\alpha_{F}}$
insertion (the other terms will either contain some trace of tensor
$N^{\alpha_{1},...,\alpha_{F}}$ or will be
proportional to  some power of $v$ and will disappear in the Heisenberg
fixed point). The graphs for $\Gamma^{(F)}$ are obtained from the
usual four-point graphs (see fig. \ref{fig:graphs})
by replacing each four-point vertex in turn by
$v N^{\alpha_{1},...,\alpha_{F}}$. In three loop approximation which
we are going to consider here  there appear two more graphs in
$\Gamma^{(F)}$ which can not be produced in this manner (they are
shown in fig. \ref{fig:stargr}).
Finally the expressions for $\beta$- and $\gamma$-functions read
\footnote{We changed the scale for the renormalized couplings
$u = u^{'}(n+8)D/6$, $v = v^{'}D/6$  and beta functions
$\beta_{u} (u)= 6\beta_{u^{'}} (u^{'})/[(n+8)D]$,
$\beta_{v} (u)= 6\beta_{v^{'}} (u^{'})/D$,
($D$
being the one-loop integral:
$D = \frac{1}{(2\pi)^d}\int \frac {d\vec{k}}{(k^2+1)^2}$,
$u^{'}, v^{'}$ being the renormalized couplings corresponding to the bare
couplings $u_{0}, v_{0}$),
to define a convenient numerical scale in which the first  two
coefficients of $\beta_{u} (u)$ are -1 and 1 at $d=3$.}:
\bee
\beta_{u} &=& - (4-d) u \left [1 - u + \beta _{u}^{2LA} u^2
+ \beta _{u}^{3LA} u^3 + \dots \right ],
\label{eq:rg2}\\
\beta_{v} &=& - (4-d) v  [\frac{\delta_{F}}{(4-d)}
- \frac{F(F-1)}{n+8}(u + \beta _{v}^{2LA} u^2 +
\nonumber\\
&&\beta _{v}^{3LA} u^3 + \dots],
\label{eq:rg3}\\
\gamma_\phi &=& - (4-d) \frac{2(n+2)}{(n+8)^2} u^2 \left [ 2 i_2 +
(4i_2 - 3i_8) u + \dots \right ],
\label{eq:rg4}\\
\tilde{\gamma}_{\phi^2} &=& (4-d) \frac{n+2}{n+8} u \left [1 +
\gamma^{2LA} u + \gamma^{3LA} u^2 + \dots \right ],
\label{eq:rg5}
\eee
here $\delta_{F}$ is the engineering dimension of the coupling $v$:
$$
\delta_{F}= F + d - \frac {Fd}{2}
$$
and expressions for two-loop ($\beta_{u}^{(2LA)}$,
$\beta_{v}^{(2LA)}$, $\gamma^{(2LA)}$) and three-loop
($\beta_{u}^{(3LA)}$, $\beta_{v}^{(3LA)}$, $\gamma^{(3LA)}$)
contributions read:
\bee
\beta_{u}^{(2LA)} &=& \frac  {8} {(n+8)^{2}} [(5n+22) (i_{1} -
      \frac {1}{2}) + i_{2}(n+2)] ,
\nonumber\\
\beta_{v}^{(3LA)}&=& -(2i_{1}(n+4F-2)+2i_{2}(n+2)/(F-1) +
\nonumber\\
&&(-n-4F+2))/(n+8),
\nonumber\\
\gamma^{(2LA)} &=& \frac{6}{(n+8)} (1-2i_1) ,
\nonumber\\
\beta_{u}^{(3LA)} &=& \frac{1}{(n+8)^{2}} \sum_{j=0}^{8}
      b_{u}^{j}i_{j}, \label{eq:rg9}\\
\beta_{v}^{(3LA)} &=& \frac{1}{(n+8)^{2}} \sum_{j=0}^{8}
     b_{v}^{j}i_{j},
\label{eq:rg10}\\
\gamma^{(3LA)} &=& \frac{1}{(n+8)^2} [10(n+8) - (44n+280)i_1 +
 \nonumber\\
&& + (8-3d)(n+2)i_2 +  12(n+2)i_3 +
 \nonumber\\
&& 24(n+8)i_4 + 6(n+8)i_5 + 18(n+2)i_6].
\nonumber\\
\eee
In (\ref{eq:rg9}), (\ref{eq:rg10}) $i_{0}\equiv1$. And for the
coefficients $b_{u}^{j}$,$b_{v}^{j}$  one has:
\bee
      b_{u}^{0}&=&-8(4n^{2}+61n+178),\nonumber\\
      b_{u}^{1}&=&4(31n^{2}+430n+1240),\nonumber\\
      b_{u}^{2}&=&(3dn^{2}+30dn+48d+8n^{2}+80n+128),\nonumber\\
      b_{u}^{3}&=& -12(n^{2}+10n+16),\nonumber\\
      b_{u}^{4}&=& -48(n^{2}+20n+60),\nonumber\\
      b_{u}^{5}&=&-24(2n^{2}+21n+58),\nonumber\\
      b_{u}^{6}&=&-6(3n^{2}+22n+56),\nonumber\\
      b_{u}^{7}&=&-24(5n+22),\nonumber\\
      b_{u}^{8}&=&-12(n^{2}+10n+16);
\nonumber\\
\vspace{2ex}
      b_{v}^{0}&=&(n^{2}+8nF+6n+20F^{2}-28F+56),\nonumber\\
  b_{v}^{1}&=&-4(n^{2}+7nF+5n+18F^{2}-28F+54),\nonumber\\
  b_{v}^{2}&=&-(3dnF-3dn+6dF-6d+4n^{2}-\nonumber\\
&&  8nF+48n-16F+80)/(F-1),\nonumber\\
  b_{v}^{3}&=&12(n+2),\nonumber\\
  b_{v}^{4}&=&12(nF+2n+4F^{2}-10F+20),\nonumber\\
  b_{v}^{5}&=&3(n^{2}+4nF-2n+4F^{2}+12F-24),\nonumber\\
  b_{v}^{6}&=&6(n+2F^{2}-10F+18),\nonumber\\
  b_{v}^{7}&=&4(nF-2n+14F-28),\nonumber\\
  b_{v}^{8}&=&3(n^{2}+10n+16)/(F-1).
\nonumber\\
\eee
here $i_{1} - i_{8}$ are the integrals originating from the
corresponding two- and three-loop Feynman graphs. Their numerical
values at $d=3$ are as follows \cite{NicMeiBak77,HolKro94}:
\bee
i_{1}= 2/3;\quad i_{2}=-2/27;\quad i_{3}=-.0376820725;
\nonumber\\
i_{4}= .3835760966; \quad i_{5}= .5194312413; \quad i_{6}= 1/2;
\nonumber\\
i_{7}=  .1739006107; \quad i_{8}=-.0946514319.        \label{eq:rg14}
\eee
Correspondence between $i_{1} - i_{8}$ and the
appropriate Feynman graphs is given in the Appendix A. For the
additional graphs appearing in $\Gamma^{(F)}$ one finds the
corresponding normalized numerical values to be equal:  $1$ (fig.
\ref{fig:stargr}a) and $i_{1}$ (fig.\ref{fig:stargr}b).

In the case $d=3$ expression for $\beta_{u}$ and gamma-functions
coincide with three-loop parts of appropriate expressions obtained
in \cite{BakNicMei78} and at arbitrary value of $d$ they were
given in \cite{Hol93}.

Note that expressions (\ref{eq:rg2}) - (\ref{eq:rg5}) are written
for the arbitrary number of field components $n$ and thus contain more
information that is necessary for the polymer case $n=0$.
This case was considered by Wallace and Zia \cite{WalZia75} in
three loop $\varepsilon$-expansion motivated by the question of
relevance of the $v$-coupling in the renormalization group sense.
These authors introduced a series of critical exponents $\alpha_{F}$:
the value of the critical exponent $\alpha_{F}$
is connected with the anomalous dimension $x'_F$ of $v$ at
the $O(n)$-symmetrical fixed point:
\bee
&&(2-\eta)\alpha_{F} = x'_{F}-\frac{F}{2}\eta ,\nonumber\\
&& x'_{F}=\frac{\partial\beta_{v}(u,v)}{\partial v}
\mid_{u=u^{*},v=0} - \delta_F.      \label{eq:rg15}
\eee
The relation to the star exponents is given in the limit $n=0$ by
\be \label{eq:rg16}
\gamma_{F}-1= -\nu (2-\eta) \alpha_{F} + [\nu(2-\eta)-1] F.
\ee
Thus the expressions (\ref{eq:rg2})-(\ref{eq:rg5}) may be used to
study the stability of the $O(n)$-symmetric fixed point to the
perturbation introduced by the traceless coupling.

%
\sect{Star exponents in three dimensions \label{sect:3d}}
We now proceed with the calculation of the star exponents.
Combining (\ref{eq:rg2})-(\ref{eq:rg5}) with (\ref{eq:rg16}) one
obtains  the following expression for the function $\gamma_{F}(u)$
($\gamma_{F}=\gamma_{F}(u^{*})$)
\footnote{
Note scaling relation (\ref{eq:3d10})
allowing one to find connection between  $\gamma_F(u)$ and functions
(\ref{eq:rg2}) - (\ref{eq:rg5}) on the basis of (\ref{eq:rg15}).
Critical exponent $\eta$ and the combination $2-\nu^{-1}-\eta$ are
given by the fixed point values of $\gamma_\phi$,
$\tilde{\gamma}_{\phi^2}$ (\ref{eq:rg4}),(\ref{eq:rg5})
}:
\bee
\gamma_{F} &=& 1 + (4-d) F [\gamma_{F}^{1LA} u + \gamma_{F}^{2LA} u^2
+ \gamma_{F}^{3LA} u^3 ] , \label{eq:3d1}\\
\gamma_{F}^{1LA} &=& \frac {n-F+3} {2(n+8)}, \label{eq:3d2}\\
\gamma_{F}^{2LA} &=& \frac {-1} {(n+8)^{2}}
     [(-nF+7n-4F^{2}+6F+10)i_{1} + \nonumber \\
&&   (dn^{2}-dnF+5dn-2dF+6d-4n^{2}+ 6nF- \nonumber \\
&&  34n + 8F^{2}-4F-44)/4],                 \label{eq:3d3} \\
\gamma_{F}^{3LA} &=& \frac{1}{(n+8)^{3}}
\sum_{j=0}^{8} \gamma_{F}^{j} i_{j},          \label{eq:3d4}\\
\gamma_{F}^{0}&=& \frac {-1}{8}(-d^{2}n^{3}+d^{2}n^{2}F- \nonumber \\
&& 7d^{2}n^{2}+4d^{2}nF - 16d^{2}n+4d^{2}F-12d^{2} + \nonumber \\
&& 8dn^{3}-10dn^{2}F+82d*n^{2}-8dnF^{2} - \nonumber \\
&&  36dnF+236dn-16dF^{2}-32dF + 208d-16n^{3}+ \nonumber \\
&& 28n^{2}F-260n^{2}+ 64nF^{2} + 72nF-1112n+ \nonumber \\
&& 80F^{3}-128F^{2}+ 400F-1504), \nonumber \\
\gamma_{F}^{1}&=& \frac {-1}{2} (dn^{2}F-13dn^{2}+ \nonumber \\
&& 4dnF^{2}+2dnF- 54dn+8dF^{2}-56d-8n^{2}F+100n^{2}- \nonumber \\
&& 44nF^{2}+604n-72F^{3}+ 152F^{2}-328F+1000), \nonumber \\
\gamma_{F}^{2}&=& \frac {-d}{2}(n^{2}-nF+5n-2F+6), \nonumber \\
\gamma_{F}^{3}&=& -6(-n^{2}+nF-5n+2F-6), \nonumber \\
\gamma_{F}^{4}&=& -6(-2n^{2}+nF^{2}+nF-22n+4F^{3} - \nonumber \\
&& -14F^{2}+30F-52), \nonumber \\
\gamma_{F}^{5}&=& \frac {-3}{2} (n^{2}F-3n^{2}+4nF^{2}-6nF - \nonumber
\\ && 18n+4F^{3}+8 F^{2}-36F-8), \nonumber \\
\gamma_{F}^{6}&=& -3 (-3n^{2}+nF-13n+2F^{3} - \nonumber \\
&& 12F^{2} + 28F - 30), \nonumber \\
\gamma_{F}^{7}&=& -2 (nF^{2} - 3nF+2n+14F^{2}-42F+28).
\nonumber
\eee
Calculating star exponents $\gamma_F$ (\ref{eq:3d1}) one should solve
first the fixed point equation:
\be \label{eq:3d6}
\beta_u(u^*) = 0
\ee
and then calculate series (\ref{eq:3d1}) at $u=u^*$ resulting in the
value of star exponent $\gamma_F$:
\be \label{eq:3d7}
\gamma_F = \gamma_F(u=u^*).
\ee
In order to be self-consistent substituting result of (\ref{eq:3d6})
into the series for $\gamma_F$ we introduce here
pseudo-$\varepsilon$-expansion (See Appendix B). The resulting series
for $\gamma_F$ (\ref{eq:b2}) (as well as (\ref{eq:rg2}) -
(\ref{eq:rg5})) is known to be asymptotic and the appropriate
resummation procedure is to be applied to obtain from it reliable
information.

The results given below were obtained by applying two different
resummation techniques to the series for $\gamma_F(\tau)$
(\ref{eq:b2}) as a function of ``pseudo-$\varepsilon$" parameter
$\tau$
\footnote{
Let us recall that star exponent $\gamma_{F}$ equals
$\gamma_{F}= \gamma_{F}(\tau=1) $.
}:
\be \label{eq:3d8}
\gamma_{F}(\tau)= \sum_{j} \tilde{\gamma}_{F}^{j}\tau^{j}.
\ee
First we tried simple Pad\'e-Borel resummation using [2/1]
Pad\'e-approximant  $\gamma_F^P(\tau)$  for analytical continuation
of the Borel transform $\gamma_F^B(\tau t)$ of $\gamma_{F}(\tau)$:
$$
\gamma_{F}^B(\tau t)= \sum_{j} \frac{\tilde{\gamma}_{F}^{j}}{j!}
(\tau t)^{j}
$$
and writing for the resummed function:
$$
\gamma_{F}^{res}(\tau)= \int_{0}^{\infty}d t
e^{-t} \gamma_{F}^P(\tau t).
$$
Then we applied resummation procedure based on the conformal mapping
transformation, mapping the domain of analyticity of
$\gamma_{F}^B(\tau)$ containing the real positive axis onto a circle
centered at the origin. Here we introduced a fit parameter ($b$)
considering instead of Borel-transform the Borel-Leroy transform
$\gamma_{F}^{B-L}(\tau)$ of $\gamma_{F}(\tau)$ defined by:
\be \label{eq:3d9}
\gamma_{F}^{B-L}(\tau)= \sum_{j}
\frac{\tilde{\gamma}_{F}^{j}}{\Gamma(j+b+1)} (\tau)^{j},
\ee
then
$$
\gamma_{F}(\tau)= \int_{0}^{\infty}d t t^b
e^{-t} \gamma_{F}^{B-L}(\tau t).
$$
Assuming that the singularity of $\gamma_{F}^{B-L}(\tau$) closest to
the origin is located at the point $(-1/a)$ we mapped the $\tau$
plane onto a circle with a mapping leaving the  origin invariant:
$$
w = \frac{(1+a\tau)^{1/2}-1}{(1+a\tau)^{1/2}+1},
$$
$$
\tau= \frac{4}{a} \frac{w}{(1-w)^2}
$$
and thus obtained an expression for $\gamma_F^{B-L}(\tau)$ convergent
in the whole cut plane and, as a result, the expression for the resummed
function $\gamma_{F}^{res}$. While doing this in order to weaken
a possible singularity on $w$-plane we multiplied the corresponding
expression by $(1-w)^{\alpha}$ and thus introduced one more parameter
$\alpha$. In the resummation procedure the value of $a$ (the location of
the closest singularity in pseudo-$\varepsilon$-expansion) was taken from
the known large-order behavior of $\varepsilon$-expansion series for
critical exponents \cite{BreLeGZin77}, while $\alpha$ was chosen as a
fit parameter defined by the condition of minimal difference between
resummed two- and three-loop results. The resummation procedure was
quite insensitive to the parameter $b$ introduced by the Borel-Leroy
transformation (\ref{eq:3d9}).

The results obtained are given in table \ref{tabl:exp}.  First we
give the values of the exponent $\gamma_F$ obtained by Pad\'e-Borel
resummation: the second column contains the value of $\gamma_F$ obtained
directly from resummation of the series (\ref{eq:3d8}), while the
third column gives $\gamma_F$ obtained on the base of the resummed series
for the exponent $x^{'}_{F}$ (namely such a way of calculation of
$\gamma_F$ was chosen in \cite{SchaFer92,DuplFer92} using the
$\varepsilon$-expansion method). In this case the numerical value of
$\gamma_F$ was obtained on the base of the numerical value of $x^{'}_{F}$
via scaling relation:
\be \label{eq:3d10}
\gamma_{F}= 1-\nu x^{'}_{F}+[\nu(2-\eta/2)-1]F
\ee
substituting well-known values of the exponents $\nu(d=3)=0.588$,
$\eta(d=3)=0.027$. Columns 4 and 5 give $\gamma_F$ obtained by the
resummation based on conformal mapping technique: resummation of the
series  $\gamma_F(\tau)$ (fourth column) and resummation of the
series $x^{'}_{F}(\tau)$ (fifth column).
The next columns give the results obtained by the
$\varepsilon^{3}$-expansion based on:  simple Pad\`{e} approximation
(the 6th column) and Pad\`{e}-Borel analysis neglecting or exploiting
exact results for $d=2$ (the 7th and 8th columns respectively)
\cite{SchaFer92,DuplFer92}. The last column contains Monte-Carlo data
\cite{BatKre89,BarTre87}.  For low number of arms $F\leq 5$ the
results of the different approaches agree reasonably well and are also
close to the values obtained by MC simulation. We have used two
different renormalization schemes as well as different procedures for
the resummation of the resulting asymptotic series.  Table
\ref{tabl:exp} gives thus a test for the stability of the results
under changes of the calculational scheme.  Obviously for higher
number $F>5$ of arms coincidence of the results is no longer good. The
main reason for this is that calculating the exponents combinatorial
factors lead to an expansion in $F\varepsilon$ for the $\varepsilon$
expansion and of $Fg$ when directly expanding in a renormalized
coupling $g$. For such large values of the expansion parameter even
resummation of the series fails. For large numbers of arms other
approaches to the theory of polymer stars like a self consistent field
approximation might be more useful.  We conclude that the Parisi
method of massive renormalization in fixed dimension as it is widely
applied in the theory of critical phenomena used together with an
appropriate resummation scheme is a powerful tool also for the
calculation of exponents in polymer theory in the present case leading
to a good test of previous results and methods.

%
\section*{Acknowledgements}

We would like to thank Prof. Lothar Sch\"{a}fer for valuable
discussions and advice.  This work has been supported in part by the
Deutsche Forschungsgemeinschaft, Sonderforschungsbereich ``Unordnung
und gro\ss{}e Fluktuationen" and by the Ukrainian State Committee for
Science and Technology, project No 2.3/665.
%

%
\begin{appendix}
\section{Appendix \label{sect:a}}
\setcounter{equation}{0}
Here we give the correspondence between loop integrals entering
renormalization group functions ((\ref{eq:rg2}) - (\ref{eq:rg5}) and
below) and appropriate Feynman graphs for one-particle irreducible
vertex functions. Fig. \ref{fig:graphs} give graphs up to the
three loop order for the functions $\Gamma^{(2)}$, $\Gamma^{(4)}$ (we
keep labeling of \cite{NicMeiBak77}). Two additional graphs appearing
in $\Gamma^{(F)}$ are given by fig.\ref{fig:stargr}a,b. In table
\ref{tabl:graphs} we show the correspondence between numerical values
of the loop integrals and appropriate Feynman graphs.

\renewcommand{\theequation}{B.\arabic{equation}}
\section{Appendix \label{sect:b}}

In the frames of pseudo-$\varepsilon$-expansion
\footnote{
First introduced by B.Nickel: see Ref.19 in \cite{LeGZin80}.
}
an auxiliary parameter (let us denote it by $\tau$)  is introduced to
keep track of the orders of perturbation theory. The corresponding
expression for function $\beta_u$ (\ref{eq:rg2}) reads:
$$
\beta_{u}(u, \tau) =-\tau u +\sum_{j=2}^{\infty}
\beta_{u}^{(j-1)LA}u^{j},
$$
$\beta_{u}^{jLA}$ being $j$-loop contributions.
The equality holds:
$$
\beta_{u}(u, \tau=1) = \beta_{u}(u).
$$
Now one can obtain the value of the fixed point $u^{*}$ as a series in
$\tau$
\footnote{
In fact powers of $\tau$ in a certain term of perturbation
theory correspond to the number of loops in the loop integrals
and this enables one to separate contributions from different orders
of the perturbation theory while substituting results of
(\ref{eq:3d6}) into (\ref{eq:3d7}).
}:
\be
\label{eq:b1}
u^{*}(\tau) = \sum_{j}u^{*,(j)}\tau^{j}.
\ee
This series can be substituted into (\ref{eq:3d1}) resulting in the
expression for $\gamma_F(\tau)$ being series in powers of
``pseudo-$\varepsilon$" parameter $\tau$.
and the final formula for critical exponent $\gamma_{F}$ read:
$$
\gamma_{F} = \gamma_{F}^{res}(u^{*}, \tau=1).
$$
Where $\gamma_{F}^{res}$ means the resummed (with respect to $t$)
series $\gamma_F(\tau)$. Performing expansion (\ref{eq:b1}) on the
base of the expression for the $\beta$-function (\ref{eq:rg2}) and
substituting the series for the Heisenberg fixed point $u^{*} \neq 0$,
$v^{*}=0$ into (\ref{eq:3d1}) one finally obtains the following
expansion of $\gamma_{F}$ in powers of the ``pseudo-$\varepsilon$"
parameter $\tau$:
\be \label{eq:b2}
\gamma_{F}(t)= \sum_{j=0}^{3} \tilde{\gamma}_{F}^{j}t^{j}.
\ee
The coefficients $\tilde{\gamma}_{F}^{j}$ for $j=0-3$ read:
\bee
\tilde{\gamma}_{F}^{0} &=& 1,                         \nonumber\\
\tilde{\gamma}_{F}^{1} &=& (4-d)F \gamma_{F}^{(1LA)} , \nonumber\\
\tilde{\gamma}_{F}^{2} &=&  (4-d)F \gamma_{F}^{2LA} +
\frac {8(4-d)F\gamma_{F}^{1LA}} {(n+8)^{2}} \times
\nonumber \\
&& ((5n + 22)(i_{1} - \frac {1}{2}) + (n+2)i_{2}),  \nonumber \\
\tilde{\gamma}_{F}^{3} &=&  (4-d)F \gamma_{F}^{3LA} + \frac
     {16(4-d)F\gamma_{F}^{2LA}}{(n+8)^{2}} [(5n+22)(i_{1} -\frac{1}{2}) +
     \nonumber \\
&&    (n+2)i_{2}] + (4 - d)F \gamma_{F}^{1LA} [\frac {128}{(n+8)^{4}}
       ((5n+22)(i_{1}-\frac{1}{2}) +
\nonumber \\
&&  +(n+2)i_{2})^{2} + \frac {1}{(n+8)^3} (-32n^{2} -
 488n -1424 + 4(31n^{2}+
\nonumber \\
&& 430n+ 1240) i_{1} +  (3d +8)(n+2)(n+8)i_{2} -
\nonumber \\
&& 12(n+2)(n+8)i_{3} - 48(n^{2} + 20n + 60)i_{4} -
\nonumber \\
&&  24(2n^{2} +21n +58)i_{5} - 6(3n^{2} + 22n + 56)i_{6} -
\nonumber \\
&& 24(5n+22)i_{7} - 12(n+2)(n+8)i_{8})].              \nonumber
\eee
expressions for $\gamma_{F}^{1LA}$, $\gamma_{F}^{2LA}$,
$\gamma_{F}^{3LA}$ are defined in (\ref{eq:3d2}) - (\ref{eq:3d4}).
\end{appendix}
\newpage
\begin{table}[H]
\label{tabl:exp}
\caption {Values of the star exponent $\gamma_{F}$ obtained in
three-dimensional theory  (columns 2,3,4,5) in comparison with the
results of $\varepsilon$-expansion (columns 6,7,8) and Monte-Carlo
simulations (column 9). See the text for a full description.}
\begin{centering}
 \begin{tabular}{rrrrrrrrl}
\\
\\
 \hline\\
 $F$ & \multicolumn{2}{c} {$d=3$} & \multicolumn{2}{c}{$d=3$} &
\multicolumn {3}{c} {$\varepsilon$-expansion } & MC   \\
& \multicolumn{2}{c} {Pad\'e-Borel} & \multicolumn{2}{c}
{conf.mapping} &
\multicolumn {3}{c} {\cite{SchaFer92,DuplFer92} } &
 \cite{BatKre89,BarTre87}  \\
 \\
 \hline
\\
 3 &  1.06 &  1.05 & 1.06 &1.06 &  1.05 &  1.05 &  1.07 &  1.09 \\
 4 &  0.86 &  0.86 & .86  &.83  &  0.84 &  0.83 &  0.85 &  0.88 \\
 5 &  0.61 &  0.61 & .58  &.56  &  0.53 &  0.52 &  0.55 &  0.57 \\
 6 &  0.32 &  0.32 & .24  &.22  &  0.14 &  0.18 &       &  0.16 \\
 7 & -0.02 & -0.01 & -.17 &-.17 & -0.33 & -0.20 &       &        \\
 8 & -0.40 & -0.36 & -.63 &-.62 & -0.88 & -0.60 &       & (-0.99, -0.30) \\
 9 & -0.80 & -0.72 & -1.14&-1.11& -1.51 & -1.01 &       &       \\
 \\
 \hline
 \end{tabular}
 \end{centering}
 \end{table}

\begin{table}[H]
\label{tabl:graphs}
\caption {Normalized values of the loop integrals (for graphs 2-M1 -
5-S3 for two-point function $\Gamma^{(2)}(k)$ the corresponding
derivative
$\partial/\partial k^{2}\mid_{k^2=0}$
is given.}
\begin{centering}
\begin{tabular}{llllll}
\\
\\
 \hline\\
\\
Graph&Integral&Graph&Integral&Graph&Integral  \\
 &value& &value& &value  \\
 \hline
\\
2-U2 & 1     & 8-U4  &  $i_4$ & fig.\ref{fig:stargr}b&$i_1$ \\
3-U3 & 1     & 9-U4  &  $i_5$ &  2-M1 &  0    \\
4-U3 & $i_1$ & 10-U  &  $i_6$ &  3-S2 &  $i_2$ \\
5-U4 & 1     & 11-U4 &  $i_5$ &  4-M3 &  0    \\
6-U4 & $i_1$ & 12-U4 &  $i_7$ &  5-S3 &  $i_8$ \\
7-U4 & $i_3$ & fig.\ref{fig:stargr}a& 1 &    &  \\
\hline
\end{tabular}
\end{centering}
\end{table}
\newpage

\end{document}